# BRIGHT STARS AND RECENT STAR FORMATION IN THE IRREGULAR MAGELLANIC GALAXY NGC 2366[*]

## Authors

A. Aparicio [1], J. Cepa [1], C. Gallart [1], H. O. Castañeda [1], C. Chiosi [2], G. Bertelli [3], J.M. Mas-Hesse [4], C. Muñoz-Tuñón [1], E. Telles [5], G. Tenorio-Tagle [1], A. I. Díaz [6], L. M. García-Vargas [6], F. Garzón [1], R. M. González-Delgado [1,7], E. Pérez [7], J. M. Rodríguez-Espinosa [1], E. Terlevich [5], R. J. Terlevich [5], A. M. Varela [1], J. M. Vílchez [1]

[1] Instituto de Astrofísica de Canarias
E-38200, La Laguna, Tenerife, Canary Islands, Spain

[2] Dipartimento di Astronomia dell'Università di Padova
Vicolo dell'Osservatorio 5, I-35122, Padova, Italy

[3] Consiglio Nazionale delle Ricerche
Vicolo dell'Osservatorio 5, I-35122, Padova, Italy

[4] Laboratorio de Astrofísica Espacial y Física Fundamental
Apdo. 50727, E-28080, Madrid, Spain

[5] Royal Greenwich Observatory
Madingley Road, Cambridge CB3 0EZ, United Kingdom

[6] Dep. de Física Teórica, C-XI, Universidad Autónoma de Madrid
Cantoblanco, E-28049, Madrid, Spain

[7] Instituto de Astrofísica de Andalucía
Apdo. 3004, E-18080, Granada, Spain

**Abstract**

The stellar content of the Im galaxy NGC 2366 is discussed on the basis of CCD *BVR* photometry. The three brightest blue and red stars have been used to estimate its distance, obtaining a value of 2.9 Mpc. The spatial distribution of the young stellar population is discussed in the light of the integrated color indices and the color-magnitude diagrams of different zones of the galaxy. A generalized star formation burst seems to have taken place about 50 Myr ago. The youngest stars are preferentially formed in the South-West part of the bar, where the giant HII complex NGC 2363 is located, being younger and bluer. The bar seems to play a role favouring star formation in one of its extremes. Self-propagation however, does not seem to be triggering star formation at large scale. A small region, populated by very young stars has also been found at the East of the galaxy.

# 1. Introduction.

A considerable effort is being devoted to the analysis of the stellar content of nearby galaxies through the photometry of their resolved stars. These data, combined with spectroscopic and radio observations, provide important information about the history of the star formation and the processes of the star formation in galaxies. The understanding of these processes is relevant, since they define the path followed by the galaxy in its evolution.

In this context, late type galaxies, in particular, irregulars, are relevant for a number of reasons:

1. They are relatively simple objects, at least from the point of view that they do not show complex spiral-like structures. They are also relatively small systems. These characteristics make easier the photometric analysis of the resolved stars. It is, in principle, possible to measure virtually *all* the stars down to a certain magnitude, so having direct information about the corresponding interval of stellar masses and ages.

2. They are objects with a high activity of star formation. This means that a relatively large number of young bright and, therefore, resoluble stars is present in them.

3. They are relatively young objects, in the sense that they show usually low abundances of heavy elements and a large amount of gas. Because of this, the study of these galaxies provides information about the star formation processes in the first evolutionary steps of stellar systems.

Our project is devoted to the analysis of the stellar content in nearby galaxies. It is included in the GEFE [1] program. In this paper, we present the results of our photometric analysis of NGC 2366 (DDO 42, Markarian 71). Its coordinates are listed in Table 1, together with a summary of the properties of the galaxy. This galaxy is usually assumed to belong to the M 81–NGC 2403 group (see Sandage & Tammann, 1974), whose distance, one fourth of that to the Virgo cluster, have made the members of this group particularly interesting for the calibration of the extragalactic distance scale.

In closer relation with the present work, previous photometric data for the brightest stars of NGC 2366 have been published by Sandage and Tammann (1974) and Tikhonov et al (1991), both based on photographic photometry. These are the only photometric studies on the resolved stars in NGC 2366 carried out up to now, and no previous CCD photometry has been performed.

---

[1] The main aim of the GEFE (*Grupo de Estudios de Formación Estelar*) project is the characterization of the star formation mechanisms in galaxies. GEFE obtained 5% of the total observing time of the telescopes of the Observatories of the Instituto de Astrofísica de Canarias. This time is distributed by the *Comité Científico Internacional* among international programs.

The present paper is divided in seven sections. Section 2 presents the observations and discusses the data reduction procedures, the photometric errors and the completeness of the analysed stellar sample. The comparison with previous photographic data for this galaxy is carried out in Section 3. A description of the features appearing in the photometric color-magnitude (CM) diagrams and the cleaning of these diagrams from foreground stars are discussed in Section 4. The problem of the distance to NGC 2366 is discussed in Section 5. A discussion of the spatial distribution of stars is given in Section 6. Finally, the conclusions are summarized in Section 7.

## 2. Observations, data reduction, errors and completeness.

Observations of the stellar content of NGC 2366 were carried out in March 1992 using the 1150×1250 EEV5 chip at the prime focus of the 2.5 m Isaac Newton Telescope of the Observatory of Roque de los Muchachos in La Palma (Canary Islands, Spain). The EEV5 is a thick coated chip, with pixel size 22.5 $\mu$m, equivalent to 0".55. The total field is about $10.5 \times 11.5$ arcmin$^2$. Under very good seing conditions these parameters might lead to undersampling and compromised photometric accuracy. In our case, seeing conditions were poor (see Table 2), and problems of undersampling are not apparent. Gallart, Aparicio & Vílchez (1994b) presented a detailed discussion on this issue, and the reader is referred to their paper for more information.

Table 2 shows the journal of observations, including, in the last column, the full width at half of maximum (FWHM) of the stellar profiles, measured in each frame. *FIGARO*, *DAOPHOT* and *ALLSTAR* (Stetson 1987) were used on a VAX cluster of the Instituto de Astrofísica de Canarias to produce the photometry of the stars.

The usual procedure to transform the instrumental magnitudes derived from the Point Spread Function (PSF) fitting into aperture ones was used. We followed the same procedure described by Aparicio et al (1993) (see Stetson 1987 for a full explanation of the foundations of the method), using 12 isolated stars for $B$, 23 for $V$ and 17 for $R$. We obtained the following errors for the transformations: 0.015 in $B$, 0.018 in $V$ and 0.007 in $R$.

Atmospheric extinction corrections for each night, and instrumental to Johnson-Cousins system transforming equations for the whole campaign were then determined. To this purpose, 45 measurements of a total of 11 secondary standards of the fields of the clusters M 92 and NGC 2419 (see Christian et al 1985) were taken in each band during the three nights of the campaign. The color indices covered by the standard stars were $-0.096 < (B-V) < 1.269$ and $-0.040 < (V-R) < 0.803$. After correcting for atmospheric extinction, we derived the following set of equations transforming into the Johnson-Cousins standard system.

$$V = v + 24.033 + 0.048(B-V)$$

$$(B-V) = -0.634 + 0.929(b-v)$$

$$(V-R) = -0.363 + 0.992(v-r)$$

where capital letters are Johnson-Cousins magnitudes and small letters are instrumental magnitudes. Figure 1 shows the residuals for each transformation as a function of the magnitude (Fig. 1a) and the color indices (Figs. 1b and 1c). The corresponding zero-point errors are 0.014 in $V$ and $(B-V)$ and 0.025 in $(V-R)$. Adding quadratically this errors to those of the PSF to aperture magnitudes transformation, the final total zero point errors result: 0.02 in $V$, 0.03 in $(B-V)$ and 0.03 in $(V-R)$.

Our final photometric list is given in Table 3, available in the ApJ-AJ CDROM series. This table contains 1959 stars with photometry in $V$ and, at least, one more band. Only stars with ALLSTAR's sigma values smaller than 0.2 have been considered. Column 1 lists an identification number for each star; columns 3 and 4 give the coordinates in pixels; columns 5 to 7 respectively list the $(B-V)$ color index, the $V$ magnitude and the $(V-R)$ color index. Figure 2 shows the finding chart of the stars given in Table 3. Coordinates are given in pixels and correspond to those listed in this table. The sizes of the circles are proportional to the $V$ magnitudes. The frame has been divided into several regions for its analysis. Five regions correspond to particular fields in the main body of NGC 2366. The thick line shows the boundary between the full body adopted for the galaxy and the foreground field, of galactic stars. The division of the galaxy in five sub-fields will be used later, in Section 6, to discuss the spatial distribution of stars in NGC 2366.

The errors affecting the stellar photometry in crowded fields as well as the completeness factors, have been discussed in a number of papers (see, for example, Stetson 1987, Chiosi et al. 1989). In the present work, crowding have been analysed following the usual technique, based on artificial stars trials. The field of the CCD used is large enough to contain the whole galaxy and a companion field together. Since the crowding effects are different for the galaxy and the field, the artificial star trials have to be carried out separately or, at least, in such a way as to reach results statistically significant for both (in particular for the galaxy, that covers a smaller area) and, at the same time, to avoid overcrowding effects. For this reason, a large number of trials with few artificial stars each have been carried out. In summary, about 1000 artificial stars have been added to each frame, distributed in 38 proofs per frame. We have derived the total errors of the photometry and the completeness factors in the same way described in Aparicio et al (1993), for the galaxy and the companion field. The errors, estimated in this way, are less than 0.01 magnitudes for the three filters at magnitude 19.0. At magnitude 22 the errors for the stars in the area of the galaxy are 0.10 in $B$, 0.11 in $V$ and 0.15 in $R$. For the stars of the surrounding field, they are 0.08 in $B$, 0.04 in $V$ and 0.08 in $R$. In Figure 3, the resulting completeness factors as a function of the $B$, $V$ and $R$ magnitudes are plotted for the areas 3, 4 and 5 together (see Figure 2) and for the companion field (only for $V$). According to them, at 50% level of completeness, and for stars in the area of the galaxy, we reach magnitudes $B = 22.0$, $V = 23.0$, and $R = 22.8$. For comparison, the magnitude reached in the external companion field at the same level of completeness is $V = 23.5$.

## 3. Comparison with previous photometric data.

Sandage & Tammann (1974) and Tikhonov et al. (1991) have carried out photographic photometry of resolved stars in NGC 2366. Calibration of photographic photometry usually shows systematic effects. However, it is interesting to test whether our photometry is consistent with the scales that these authors derive from their photoelectric sequences. In Table 4 magnitudes are listed for the stars of the sequence of Sandage and Tammann (1974) that are not saturated in, at least, one of our frames. Column 1 lists the identification letter from Sandage & Tammann (1974). Columns 2 and 3 show their photometric results. Columns 4 and 5 list the photometry by Tikhonov et al. (1991). These authors give photoelectric magnitudes for two stars of the sequence of Sandage & Tammann (1974) (stars C and E). For the remainder stars, their photografic magnitudes are given. Finally, in columns 6 and 7, our results are given for the stars not saturated.

Figure 4a shows the Sandage and Tammann's (1974) $B$ and $V$ magnitudes, plotted against ours. The differences between our photometric scale and that of Sandage and Tammann (1974) lie within the dispersion of the data. However, some systematic color effect might be present. Figure 4b shows the Tikhonov et al.'s (1991) $B$ and $V$ magnitudes of the Sandage and Tammann's photoelectric sequence, plotted against ours. In both cases, the agreement is good, in particular for the magnitudes photoelectricy obtained by Tikhonov et al. (1991). But this is not the case when magnitudes of the Tikhonov et al.'s main list (their table 3) are compared with ours. The results are shown in Fig. 4c for $B$ and $V$. Large zero-points differences exist in both cases and the dispersion of the data is also apparent. The fact that photographic photometry is usually noisier than photoelectric or CCD photometry is probably the reason for this disagreement.

# 4. Photometric Diagrams.

Figures 5a and 5b show the $V$ vs. $(B-V)$ and $V$ vs. $(V-R)$ color-magnitude (CM) diagrams for the stars in the full area of the galaxy (the zone internal to the thick line in Figure 2). Contamination by foreground Milky Way's stars is expected to be important due to the low galactic latitude of NGC 2366 (see Table 1). This fact is visible in Figures 6a and 6b, where the CM diagrams of stars of the companion field (the zone external to the thick line in Figure 2) are plotted.

To decontaminate the CM diagrams of NGC 2366 from foreground stars, a procedure similar to that explained by Mateo & Hodge (1986) has been followed. This procedure is based in randomly removing stars from the CM diagram of the galaxy, but in such a way that the total number of stars removed in each area of the diagram be equal to the total number of stars in the companion field, after correcting to compensate for the different surface of galaxy and field and for the crowding effects, which are different in the galaxy and in the field. The resulting CM diagrams are shown in Figure 7a and 7b. Note, in particular, that virtually all the stars brighter than $V \simeq 19.0$, have been removed.

The CM diagrams of NGC 2366 are typical of a galaxy with active star formation. They are similar to those of other Im galaxies, eg Sextans A (Aparicio et al. 1987), NGC 3109 (Bressolin et al. 1993; Greggio et al. 1993), WLM (Ferraro et al 1989) or NGC 6822 (Gallart et al. 1994a; Gallart et al. 1994b). The diagrams of these galaxies are usually characterized by the presence of two main plumes separated by a scarcely populated region. The two plumes are not clearly visible in NGC 2366. The blending produced by the poor seeing together with the distance to the galaxy, larger than that of the Local Group members, is very likely the reason for this confussion. In any case, the region corresponding to the blue plume is placed at $(B-V)$ about –0.3 to 0.3, $(V-R)$ about –0.15 to 0.35 and the region corresponding to the red plume is placed at $(B-V)$ about 1.0 to 1.5, $(V-R)$ about 0.6 to 0.9.

At the distance of NGC 2366, the limiting magnitudes that we have reached imply that our photometry extend only to stars brighter than $M_V \simeq -4.5$, $M_{V_0} \simeq -5$ (50% completeness level) in the $V$ vs. $(V-R)$ diagram, which is that with the best completeness. As a first approximation, using data for bolometric corrections from Schmidt-Kaler (1982), it results in a luminosity of $\log(L/L_\odot) \simeq 4.2$. This means that we could barely detect some red, intermediate-mass stars of masses down to about $5 M_\odot$; but, certainly, no low or intermediate mass star of the asymptotic or red giant branches can be present in our diagram contrary to what happens in NGC 6822 (Gallart et al. 1994a, b) and Pegasus (Aparicio 1994; Aparicio, Gallart & Vílchez, 1994). The problem is more complicated for stars in the blue plume. Our limiting magnitude implies that we are observing main sequence stars down to only O9 (over $15 - 20 M_\odot$). However, most of the stars of the blue plume are probably already evolved away from the main sequence, and a number of stars in the red edge of the blue plume might be less massive stars in their blue-loop evolutionary phase. In any case, the bolometric corrections that affect this part of the CM diagram are large and they change quickly as a function of the spectral type. This complicates the determination of the spectral type of each star as well as the analysis of the blue stellar population. In Section 6, we shall discusse in more detail the age of the stellar populations in NGC 2366.

## 5. The distance to NGC 2366

The M81/NGC 2403 group, to which NGC 2366 probably belongs, is placed at a distance (3-4 Mpc), that makes it of great importance for the calibration of secondary distance indicators. A large number of papers have been published about this subject (see de Vaucouleurs 1978b, and references therein). Freedman et al. (1994) have recently published the light curves of 30 cepheids stars in M 81, obtained with the Space Telescope. They derive a distance to this galaxy of 3.63 Mpc. But few cepheid stars have been discovered and measured until now in the other galaxies of this group; in particular, none in NGC 2366.

Estimates of the true distance modulus of NGC 2366 that are found in the literature range from 27.07 to 27.65. Three of them are based on the blue supergiants: Sandage & Tammann (1974) found a value of 27.07; de Vaucouleurs (1978b) found 27.10 and Tikhonov et al (1991) found 27.65. Using the size of the HII regions, Sandage & Tammann (1974) found 27.07. Finally, Tikhonov et al (1991) found 27.62, from the brightest red supergiants. A further value was provided by Tully (1987) based on the local velocity field: using $H = 75 K m s^{-1} Mpc^{-1}$, he derived $(m - M)_0 = 27.35$.

We can provide a new estimate of the distance modulus using our data and the method of the brightest blue and red supergiants in the galaxy. The problems of this method have been discussed by several authors, and the papers by Schild & Maeder (1983) and by Greggio (1986) are very enlightening. But, till now, it is one of the few ways we have to estimate the distance to a large number of objects, NGC 2366 being among them. The average $B$ magnitude of the three brightest blue stars (using the criterion of $(B - V) \leq 0.5$) is $<B>_3 = 19.17$. The average $V$ magnitude of the three bightest red ones (with $(B - V) \geq 1.5$) is $<V> = 20.15$. Using data given by de Vaucouleurs et al (1991) for the interstelar extinction and the integrated magnitude of the galaxy and equation 5 from de Vaucouleurs (1978a), we derive a true distance modulus for NGC 2366 of $(m - M)_0 = 26.99$ from the blue stars. On the other hand, using the result by Sandage & Tammann (1974) that the average absolute magnitude for the red supergiants in a galaxy is $<M_V>_3 = -7.7$, we obtain for NGC 2366, from the red stars, $(m - M)_0 = 27.49$. Alternatively, using the fits given by Aparicio et al (1987), we obtain $(m - M)_0 = 27.01$ from the blue supergiants and $(m - M)_0 = 27.58$ from the red supergiants. In the following, we will adopt the value $(m - M)_0 = 27.3$, which is an average of the four estimates. It corresponds to 2.9 Mpc. A summary of the different estimates found in the literature is given in Table 5.

The question of the distance to this galaxy is still open, and the discovery of cepheid stars in it would be quite important. The comparison between the Tikhonov et al.'s magnitudes and ours would be very usefull to make a first list of cepheid candidates. Unfortunately, the dispersion of the data (see Figure 4c) does not allow for any step in this direction.

## 6. The spatial distribution of stars in NGC 2366

The morphology of NGC 2366, patched by a number of HII regions, indicates that star formation is ongoing in many regions of the galaxy. In particular, NGC 2366 includes a giant HII region, named NGC 2363, placed in the South-West part of the galaxy. The GEFE collaboration (González-Delgado et al. 1994) has recently carried out a detailed study of the physical conditions and chemical composition of this giant HII complex.

To test the possible differences in the star forming activity across NGC 2366, we have divided the galaxy into five regions. These are shown in Figure 2. Four regions (2, 3, 4 and 5) have been defined along the bar and a further one in South-West (region 1). Region 4 has been differenciated from region 5 because of its apparently higher current star forming activity. The integrated magnitude and color indices of these regions together with the population of their CM diagrams provide valuable information about the general distribution of the stellar population in the galaxy.

Table 6 lists the integrated $V_0$ magnitude and $(B - V)_0$ and $(V - R)_0$ color indices (columns 2 to 4) for each of the regions defined in Figure 2 (column 1). Magnitude and colour indices have been obtained from sky-subtracted calibrated frames, so that resolved and unresolved galaxy light is included in the results. These have been corrected for internal and galactic extinction using $A_B = 0.47$, $A_V = 0.35$, $A_R = 0.26$. These values have been obtained using $A_B = 0.47$ given by de Vaucouleurs et al (1991) (see Table 1) and the Cardelli et al.'s (1989) extinction law, with $R_V = 3.1$. Region 2 has been omitted because of the low signal to noise ratio.

A first look at Table 6 shows that all the regions we have defined have $(B - V)_0$ between 0.00 and 0.12. A typical low metallicity burst of star formation at ages between 2 and 5 Myr (González-Delgado et al. 1994 have derived an age of about 3-5 Myr for NGC 2363 from WR features) would show $(B - V)_0$ close to 0.0 (Leitherer & Heckman 1995). Note that the integrated colors of regions dominated by an intense star formation episode can depart significantly from the colors of the stellar population alone, due to the contribution of the nebular continuum emission. For NGC 2363, for example, this contribution can account to up to 50% (V) and 70% (R) of the continuum, as shown by Mas-Hesse & Kunth (1991). This contribution is already included in the model predictions by Leitherer & Heckman (1995). Therefore, only region 4 could be truly dominated by blue stars formed during the last burst of star formation. All the other regions show $(B-V)_0$ color indices redder than expected for a very young star formation episode. Comparing with the $(B-V)_0$ predictions of Leitherer & Heckman (1995) we see that the integrated $(B-V)_0$ of these regions are consistent with starforming episodes having taken place some 20-50 Myr ago. *It is interesting to note that older stars would yield significantly redder color indices, so that it seems that a major event of star formation took place along the whole galaxy some few tens of million of years ago.*

The results listed in Table 6 can be also compared with data given in Table 1 for the whole galaxy. From this table, $(B-V)_0^{bi} \simeq 0.46$, $(V-R)_0^{bi} = -0.12$ are obtained. $(V-R)_0^{bi}$ is too blue compared with $(B-V)_0^{bi}$. The reason for this is perhaps an inconsistency between magnitudes in the three filters, which have been obtained by different authors. If we limit to data by Pierce & Tully (1992), then $(B-R)_0^{bi} = 0.29$. Since this is the colour index integrated for the whole galaxy, it is expected to be redder than those of regions 1 to 5, as the galaxy extends further away from these regions and includes a larger proportion of unresolved stars. In fact, $(B-R)_0^{bi}$ is larger than the values corresponding to regions 1 to 4, but it is similar to that of region 5, indicating that this region is populated by a distribution of stellar masses and ages similar to that integrated for the whole galaxy.

Similar results can be found if we analyze in detail the $M_{V_0} vs (V-R)_0$ diagrams of the five regions, which are shown in Figures 8a to 8e. A mean reddening of $E(V-R) = 0.09$, which results from the adopted extinction, and a distance modulus of $(m-M)_0 = 27.3$ (see Section 5) have been used. Foreground stars have been statistically removed from these diagrams, as explained in Section 4. We have superimposed on the diagrams isochrones of Z=0.004 and Y=0.27 from Bertelli et al. (1994). Isochrones ages are 6, 10, 20, 40 and 100 Myr, from top to bottom. The diagrams show that stars are concentrated around the isochrones corresponding to about 40 Myr. Many of these stars, with $M_{V_0}$ around –5 to –6, are presently in the He burning phase. The diagrams do not provide clear information for older, fainter stars, which are not resolved. However, the age value of 40 Myr agrees with the results derived from the integrated $(B-V)$ colors.

Apart from this population of stars around 40 Myr old the CM diagrams show the presence of much younger stars. Regions 1 and 3 show the highest concentration of stars younger than 20 Myr; Region 3 contains several resolved stars younger than 6 Myr. Following the Bertelli et al.'s models, stars of this age can have up to 34 $M_\odot$, but Region 3 includes the giant HII region NGC 2363, and other younger, more massive stars have to be within this region, embeded into HII regions gass emision, and they have not been resolved. In fact, González-Delgado et al. (1994) have found WR features in NGC 2363 associated with the presence of WC stars. This implies ages between 3 and 5 Myr for the region. This has to be the case also of Region 4, whose blue integrated colors indicate the presence of a young star-forming episode. Region 2, in the North-East part of the galaxy, is the less populated by very young stars.

The integrated color indices and the structure of the CM diagrams indicate that an age gradient may be present in the recent star formation episodes in NGC 2366, in the sense that the youngest population are preferentially placed in the South-West part of the bar. This picture is common to many Im galaxies and could indicate that a self propagating mechanism is acting, triggering large scale star formation. If this is right, the velocity of the star formation propagation can be estimated. One could assume an age of 50 Myr for the latest star formation event in region 2 and 2 Myr for that of region 3. These regions are separated by about 5 arcmin, and taking into account the distance to the galaxy (see Section 5) and the inclination angle (see Table 1) we obtain about 100 Kms$^{-1}$ for the speed of propagation of stellar formation. This value seems too large when compared to the speed of propagating stellar formation ($\leq$30-50 Kms$^{-1}$ in other galaxies (see Elmegreen 1992 and references therein). Consequently, self-propagation is probably not the main mechanism triggering the star formation at large scale in NGC 2366. In fact, self-propagation is not evident in the light of the CM diagrams of the five defined regions, since young stars are present in all of them: star formation is presently taking place preferentially in regions 1, 3 and 4, but there are young stars also in regions 2 and 5. Our conclusion is that large star formation bursts are produce in the galaxy without a physical conection between them. The bar however, seems to favour the star formation in one of its extremes.

## 7. Conclusions

The stellar content of the Im galaxy NGC 2366 has been discussed, through the study of the $V$-$(B-V)$ and $V$-$(V-R)$ CM diagrams. A true distance modulus of $(m-M)_0 = 27.3$ has been obtained using the method of the brightest stars, in good agreement with most other estimates. But the determination of the distance through the light curve of cepheid stars is necesary in this galaxy, which belongs to a group of paramount importance in the calibration of secondary distance estimators.

A generalized star formation burst seems to have taken place about 50 Myr ago. It would have been strong enough to dominate the integrated color indices of most of the galaxy. The spatial distribution of the young population indicates that young stars are preferentially concentrated in the South-West part of the bar structure, in particular in the NGC 2363 giant HII complex. Older stars populate the North-East part of the bar. A small region, populated by very young stars is placed at the East of the galaxy. The possibility that self-propagation could be the trigger of star formation at large scale is unlikely. We conclude that large bursts are produced at random positions, but the galactic bar, like in other Im galaxies, seems to play a role favouring star formation in one of its extremes.

## Acknowledgements

This research has been funded by the spanish DGICyT (project reference: PB91-0531).

We acknowledge the software provided by the Starlink Project which is funded by the UK SERC.

# Figure captions

**Figure 1a:** Photometric residuals vs. magnitude of the calibration in the $V$ band.

**Figure 1b:** Photometric residuals vs. color index of the calibration in $(B-V)$.

**Figure 1c:** Photometric residuals vs. color index of the calibration in $(V-R)$.

**Figure 2:** Finding chart of the stars measured in NGC 2366 and the surrounding field. Coordinates are CCD pixels. The circles' sizes are proportional to the stars' $V$ magnitudes. The five areas in which the galaxy has been divided for the discussion of the stellar content (see section 6) are shown. The boundary of the surrounding companion field, used to correct the contamination by foreground stars, is defined as the area outside the heavy line. North is up, East is at left.

**Figure 3:** Completeness factor as a function of $B$, $V$ and $R$ magnitudes in the central part of the galaxy (zones 3, 4 and 5 of Fig.2) and the companion field (only in $V$). At 50% completeness level, $B = 22.0$, $V = 23.0$ and $R = 22.8$ are reached in the galaxy and $V = 23.5$ is reached in the companion field.

**Figure 4a:** Photoelectrical $B$ and $V$ magnitudes of the stars of the calibration sequence of Sandage & Tammann (1974) plotted against our data.

**Figure 4b:** Photoelectric and photographic $B$ and $V$ magnitudes of the stars of the calibration sequence of Tikhonov et al. (1991) plotted against our data.

**Figure 4c:** Photographic $B$ and $V$ magnitudes of the stars measured by Tikhonov et al. (1991) in NGC 2366, plotted against our data. 87 stars have been matched.

**Figure 5a:** The $V$ vs. $(B-V)$ CM diagram of the stars in the area of NGC 2366 (see Fig. 2).

**Figure 5b:** The $V$ vs. $(V-R)$ CM diagram of the stars in the area of NGC 2366 (see Fig. 2).

**Figure 6a:** The $V$ vs. $(B-V)$ CM diagram of the stars in the surrounding companion field of NGC 2366 (see Fig. 2).

**Figure 6b:** The $V$ vs. $(V-R)$ CM diagram of the stars in the surrounding companion field of NGC 2366 (see Fig. 2).

**Figure 7a:** The $V$ vs. $(B-V)$ CM diagram of the stars of NGC 2366 after correction of contamination by foreground stars.

**Figure 7b:** The $V$ vs. $(V-R)$ CM diagram of the stars of NGC 2366 after correction of contamination by foreground stars.

**Figure 8a:** The $V$ vs. $(V-R)$ CM diagrams of region 1 (see Fig. 2). From top to bottom, isochrones of $Z = 0.004$ and ages 6, 10, 20, 40 and 100 millions years are plotted.

**Figure 8b:** The $V$ vs. $(V - R)$ CM diagrams of region 2 (see Fig. 2). From top to bottom, isochrones of $Z = 0.004$ and ages 6, 10, 20, 40 and 100 millions years are plotted.

**Figure 8c:** The $V$ vs. $(V - R)$ CM diagrams of region 3 (see Fig. 2). From top to bottom, isochrones of $Z = 0.004$ and ages 6, 10, 20, 40 and 100 millions years are plotted.

**Figure 8d:** The $V$ vs. $(V - R)$ CM diagrams of region 4 (see Fig. 2). From top to bottom, isochrones of $Z = 0.004$ and ages 6, 10, 20, 40 and 100 millions years are plotted.

**Figure 8e:** The $V$ vs. $(V - R)$ CM diagrams of region 5 (see Fig. 2). From top to bottom, isochrones of $Z = 0.004$ and ages 6, 10, 20, 40 and 100 millions years are plotted.

Table captions

Table 1. Global Parameters for NGC 2366

Table 2. Journal of Observations

Table 3. Photometry

Table 4. Comparison with previous photoelectric sequences

Table 5. Distance estimates to NGC 2366

Table 6. Integrated magnitudes and colors

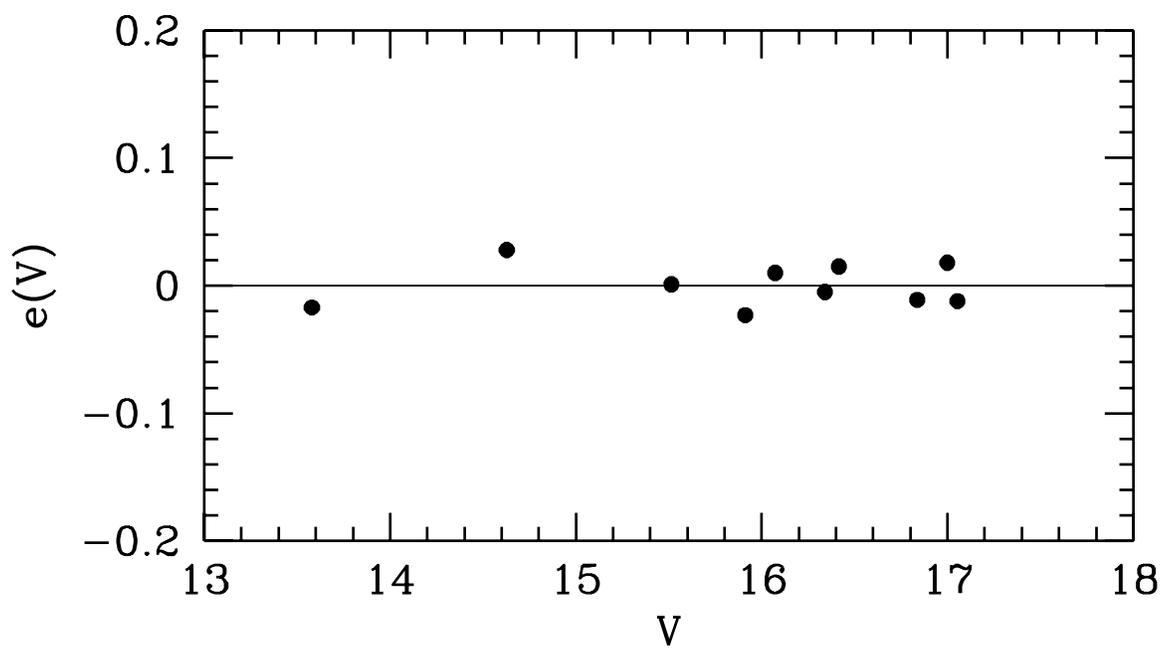

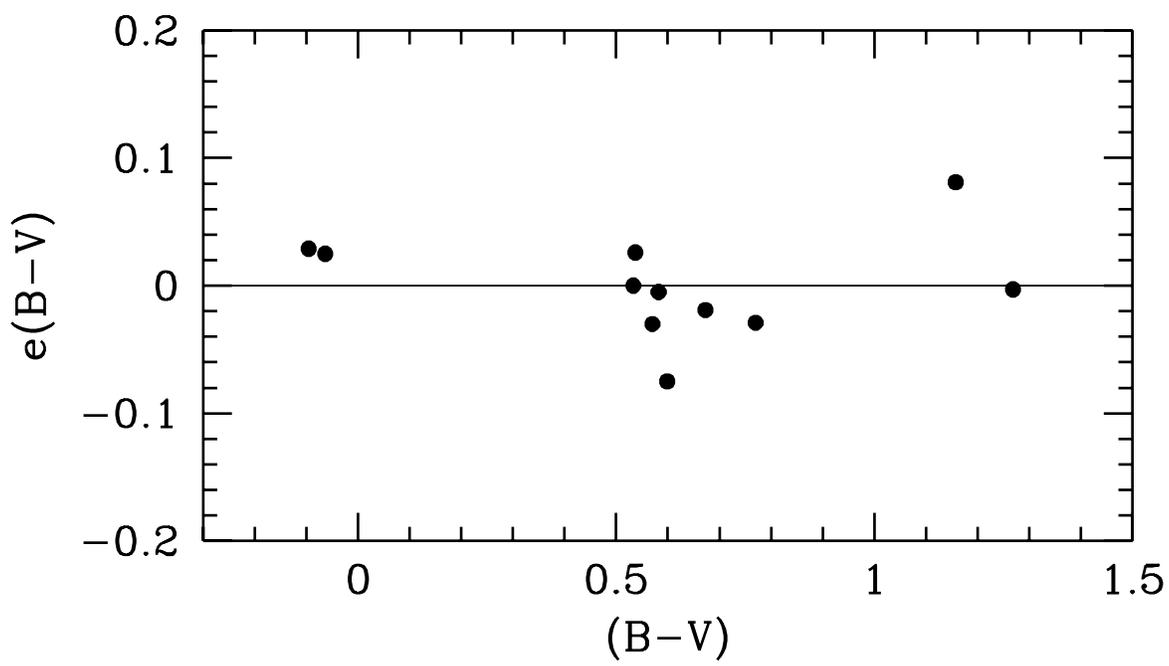

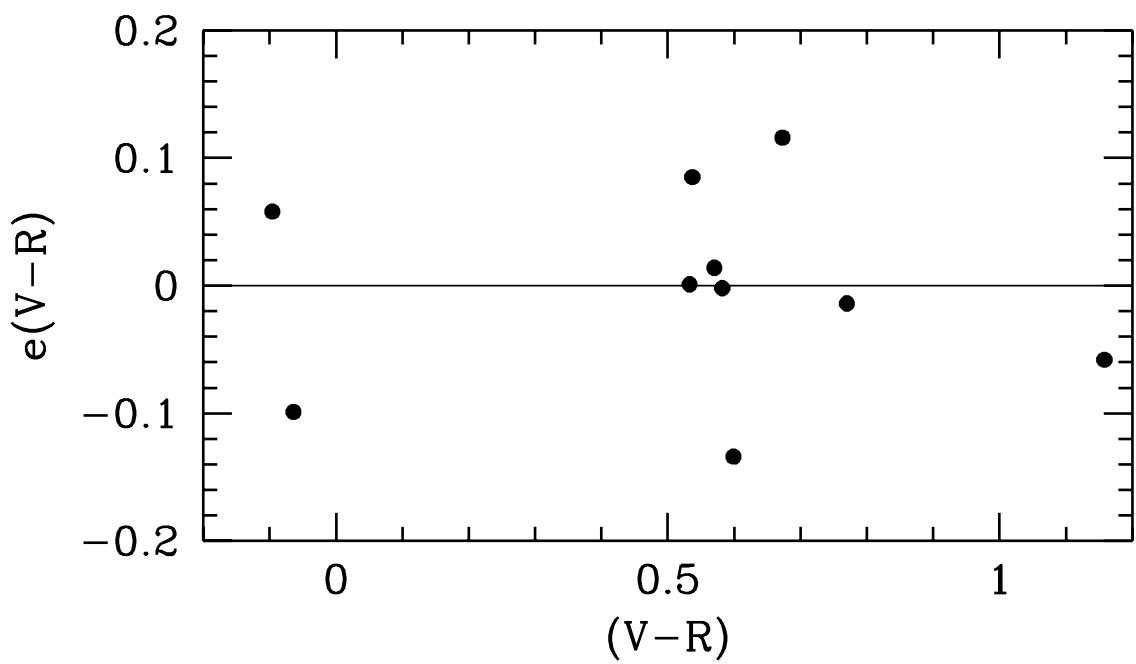

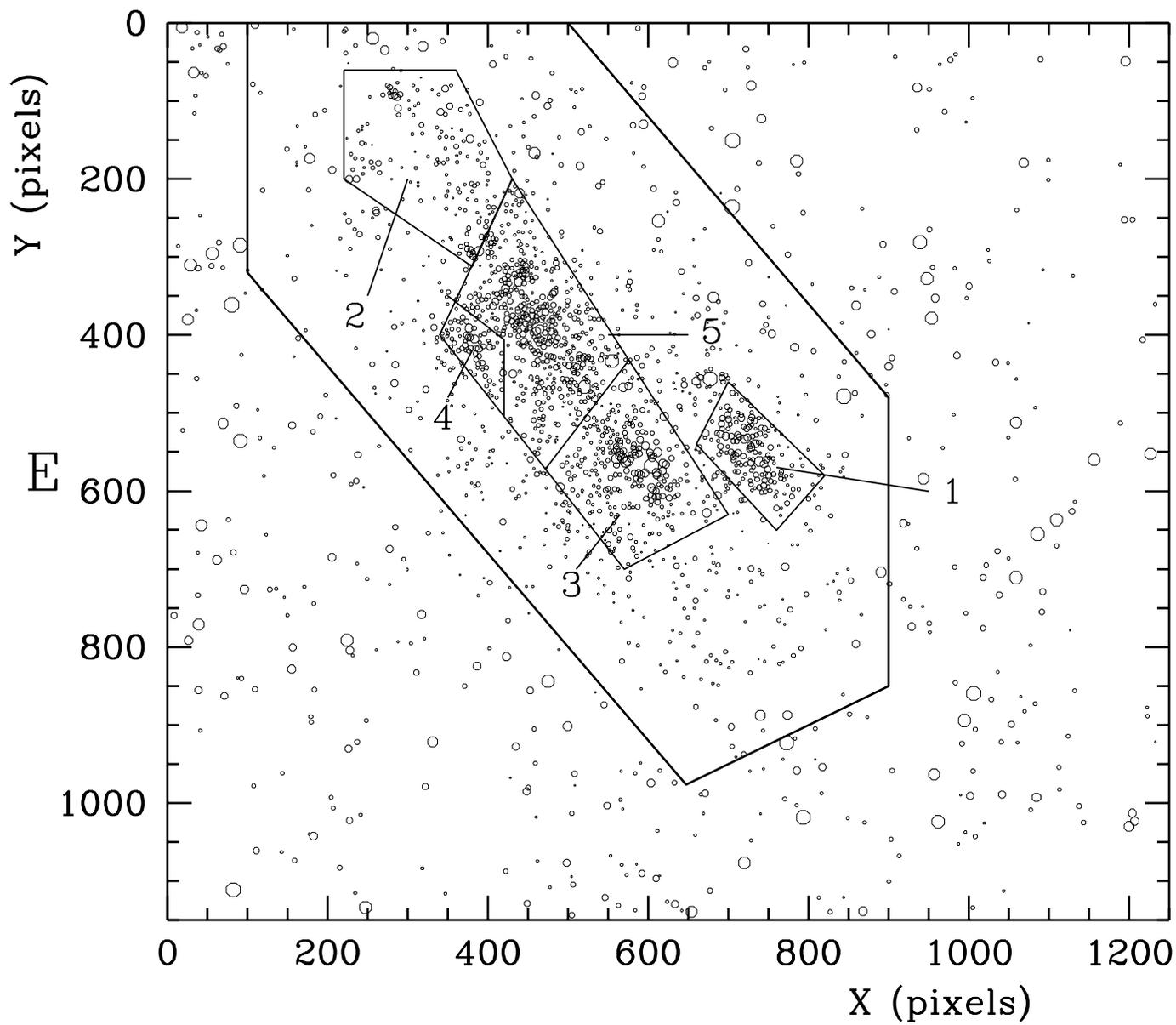

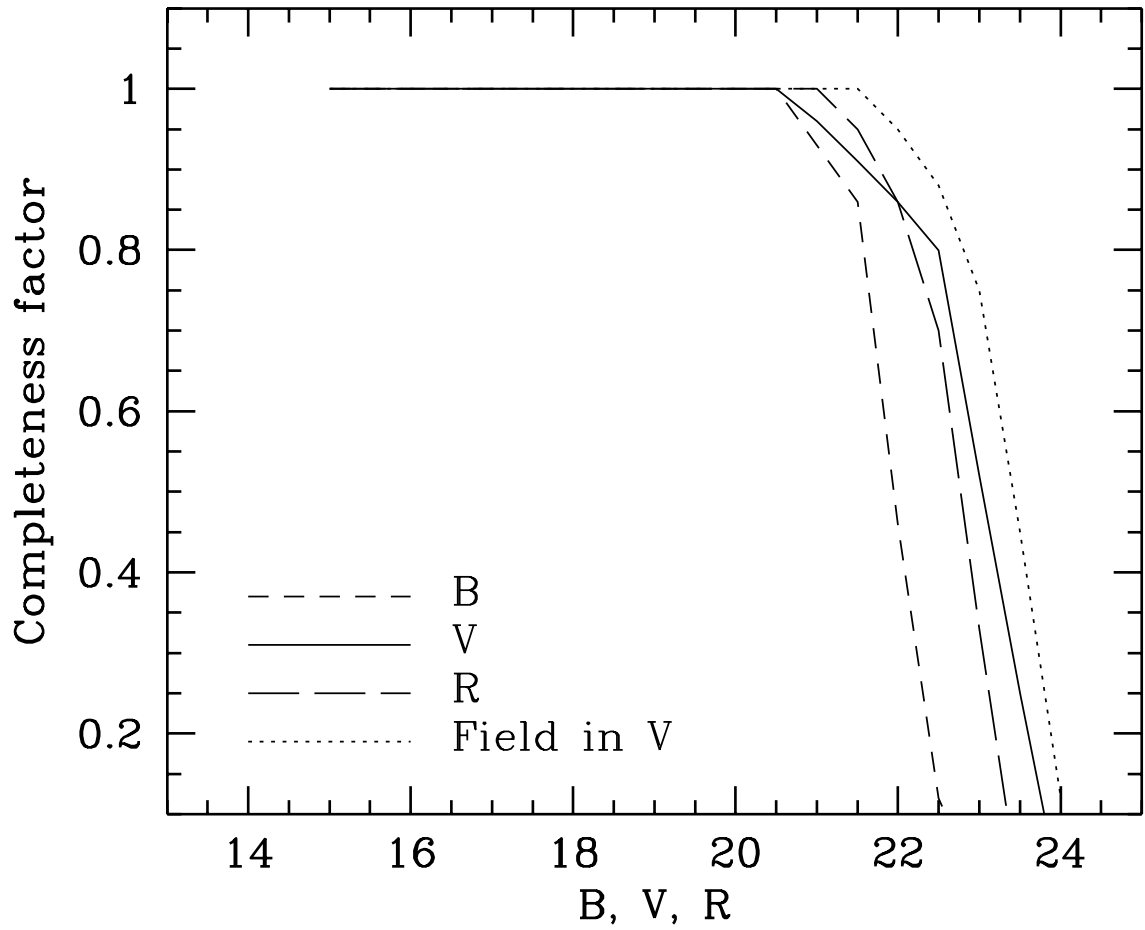

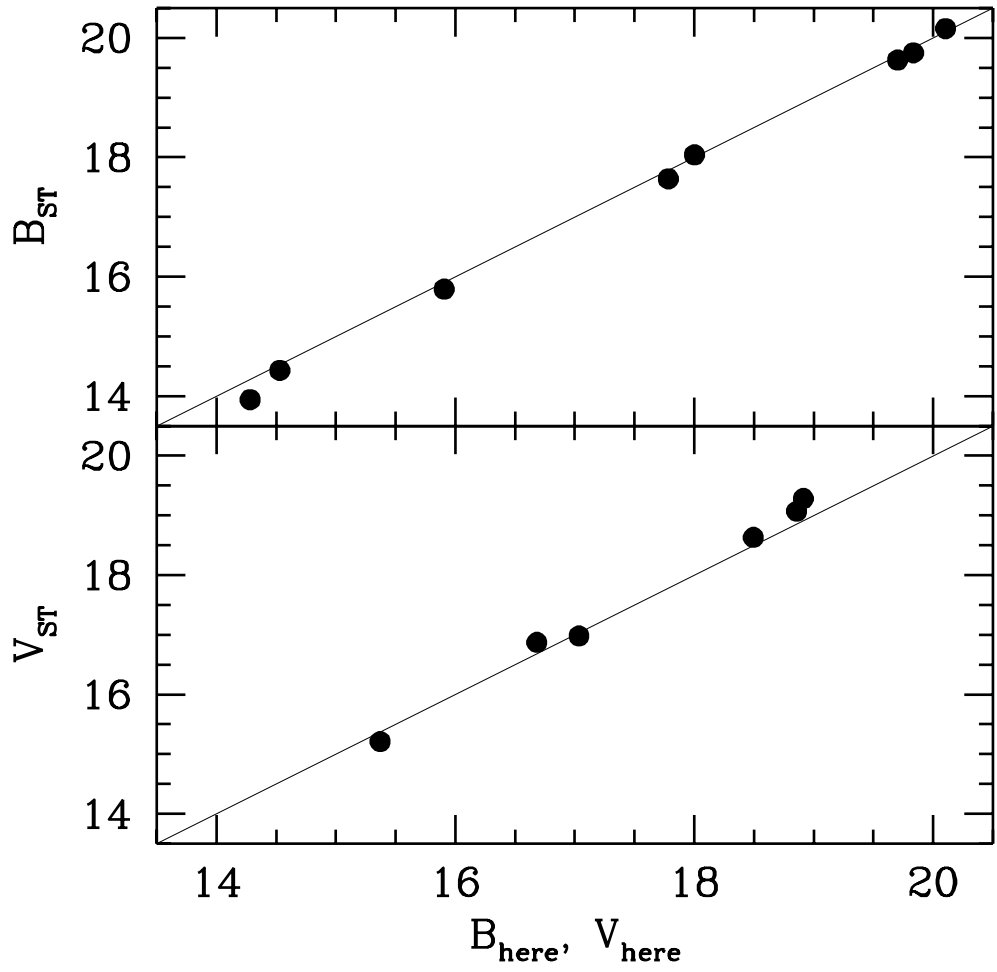

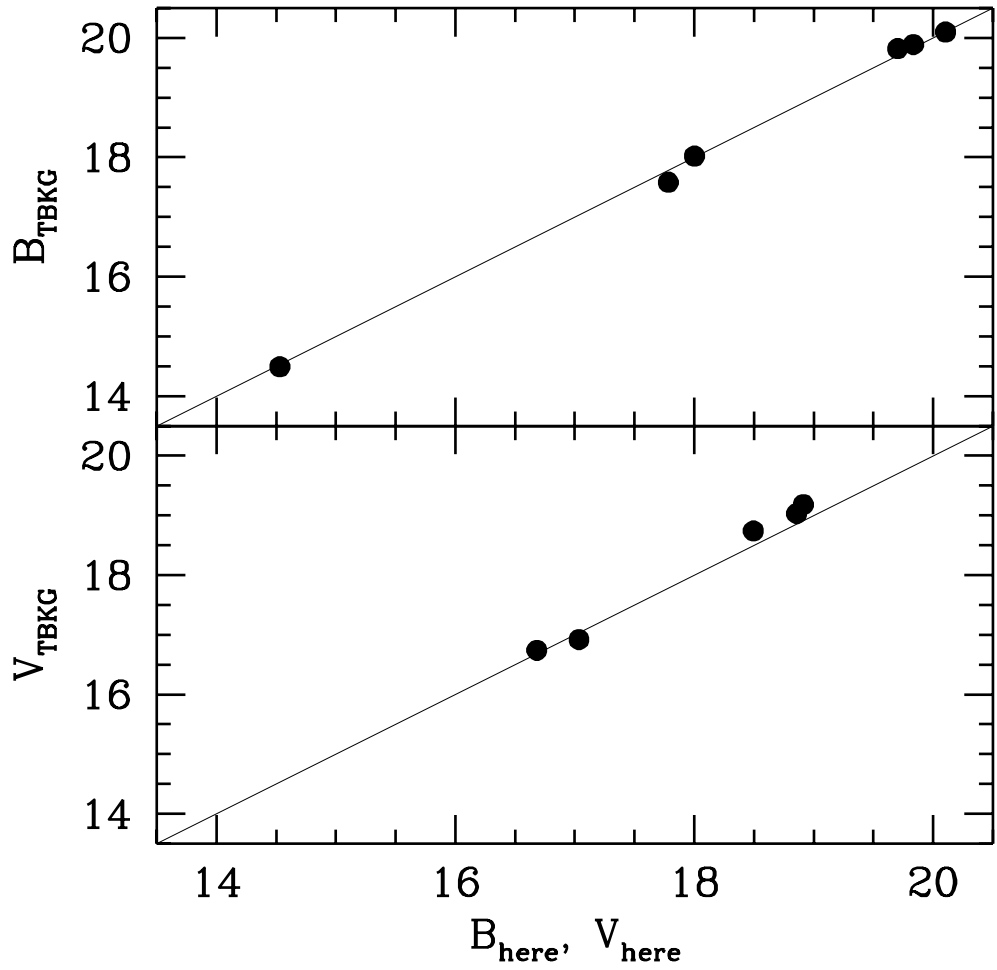

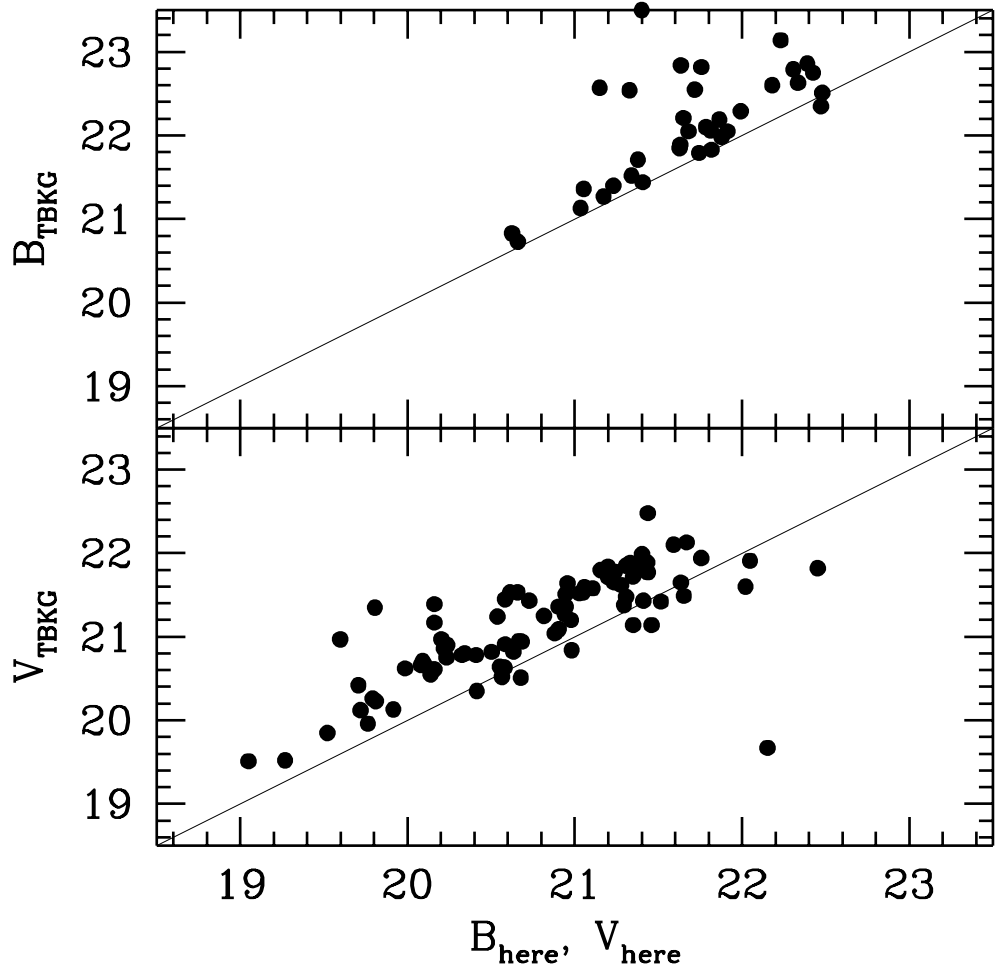

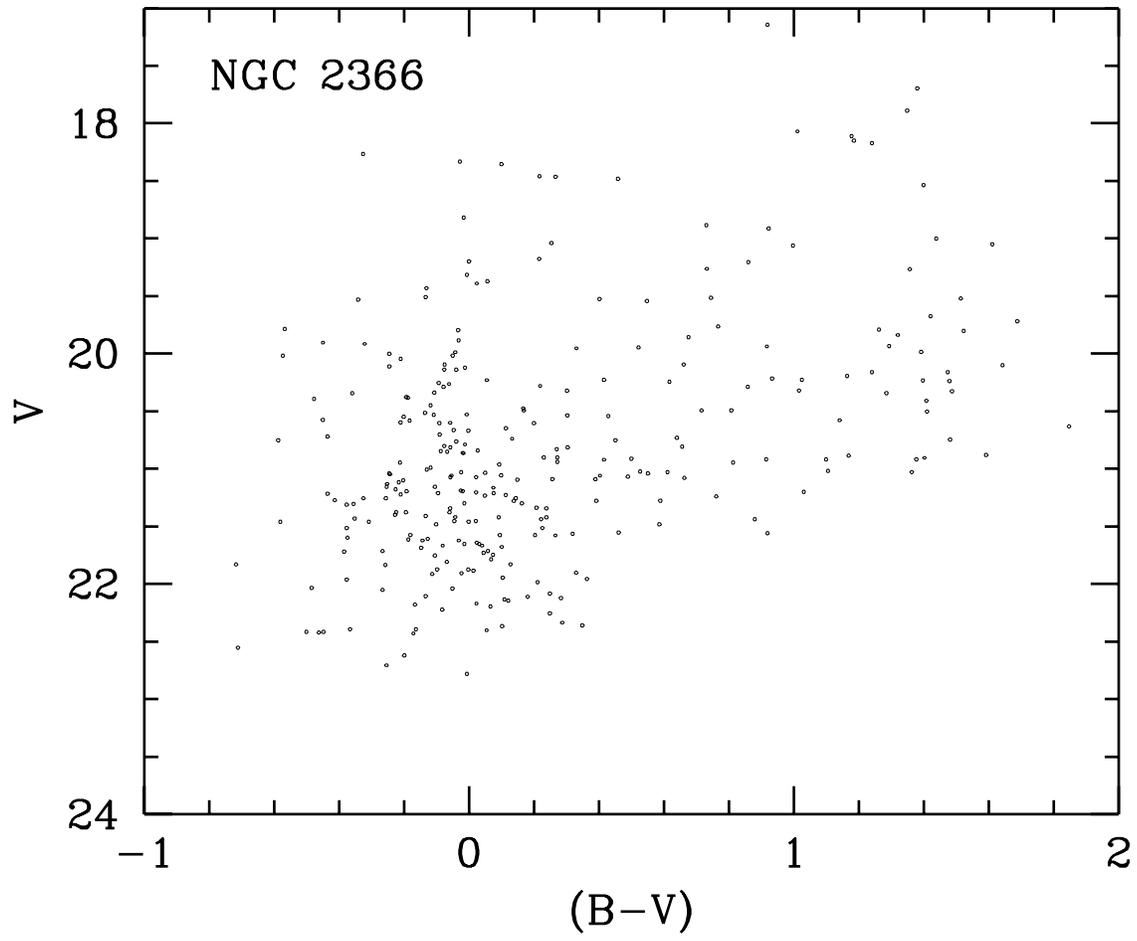

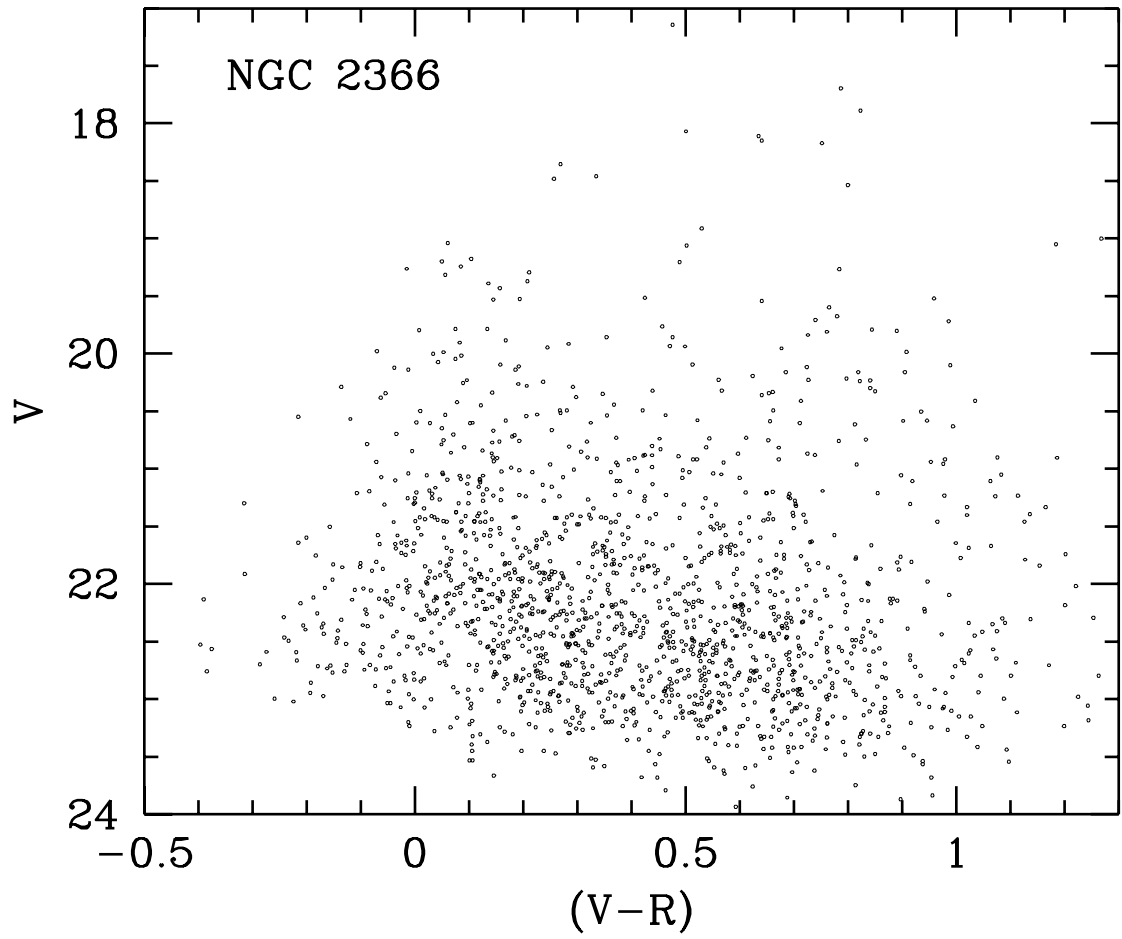

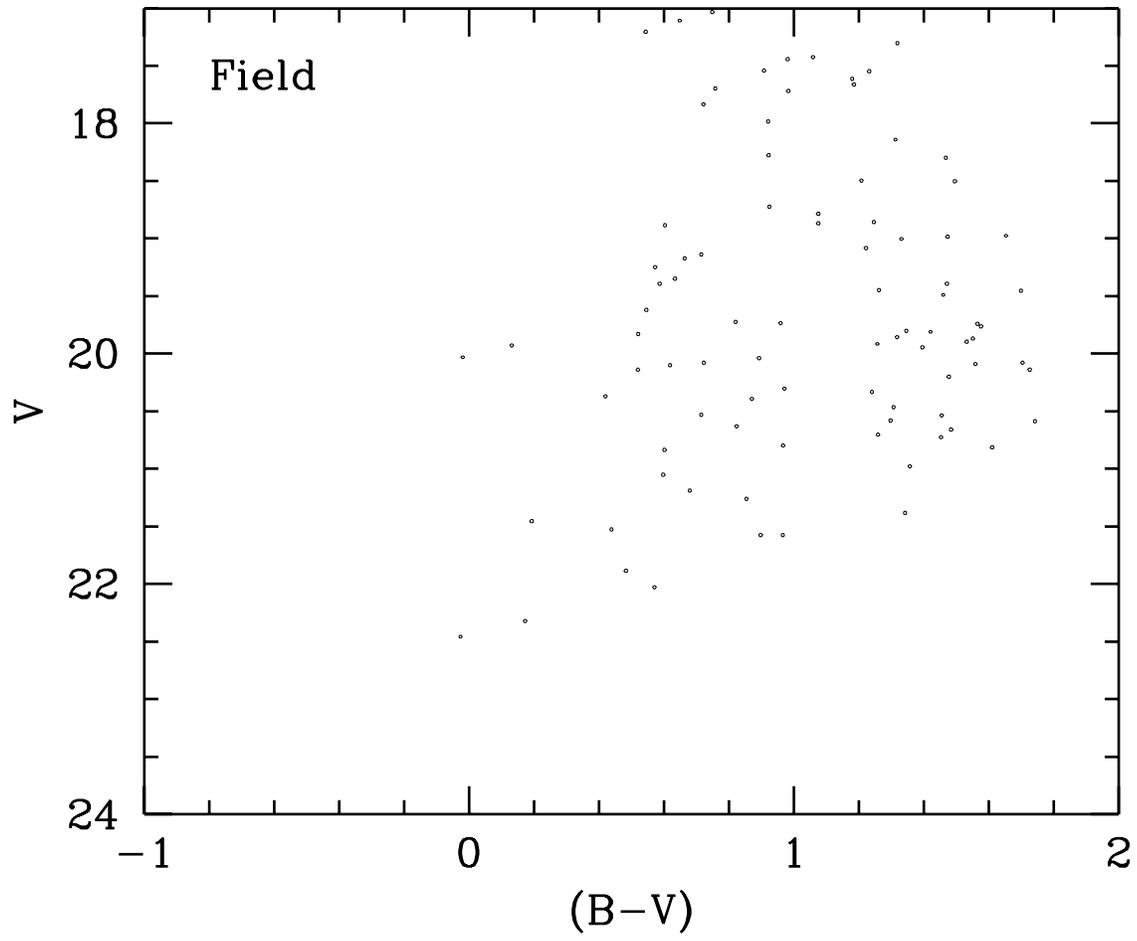

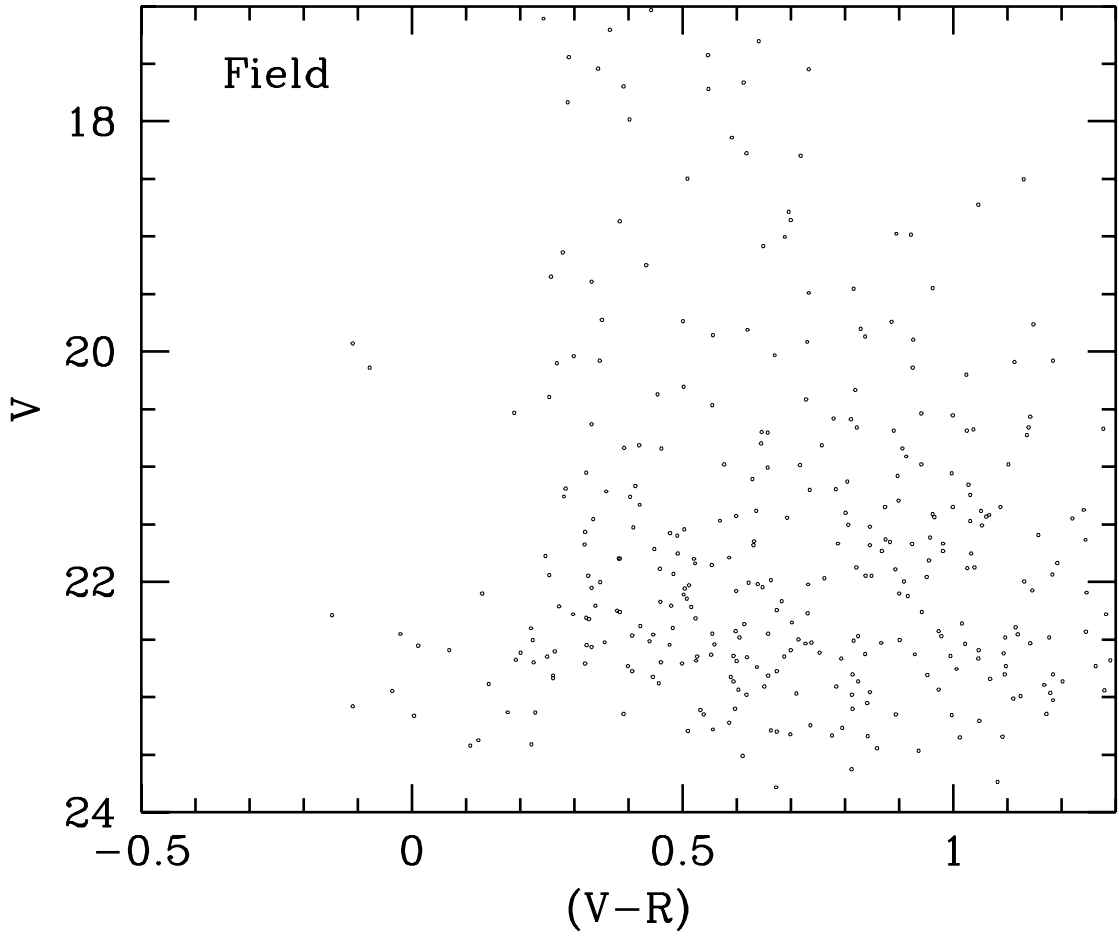

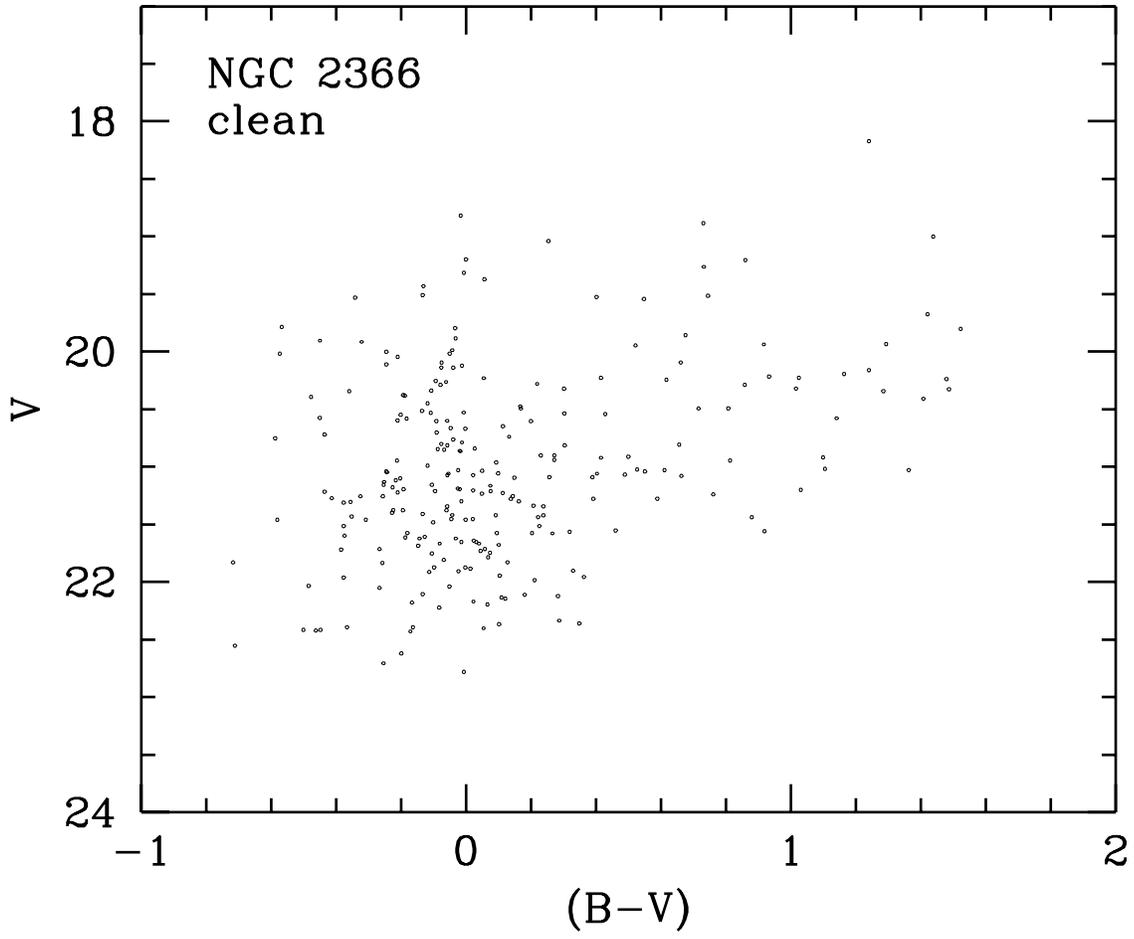

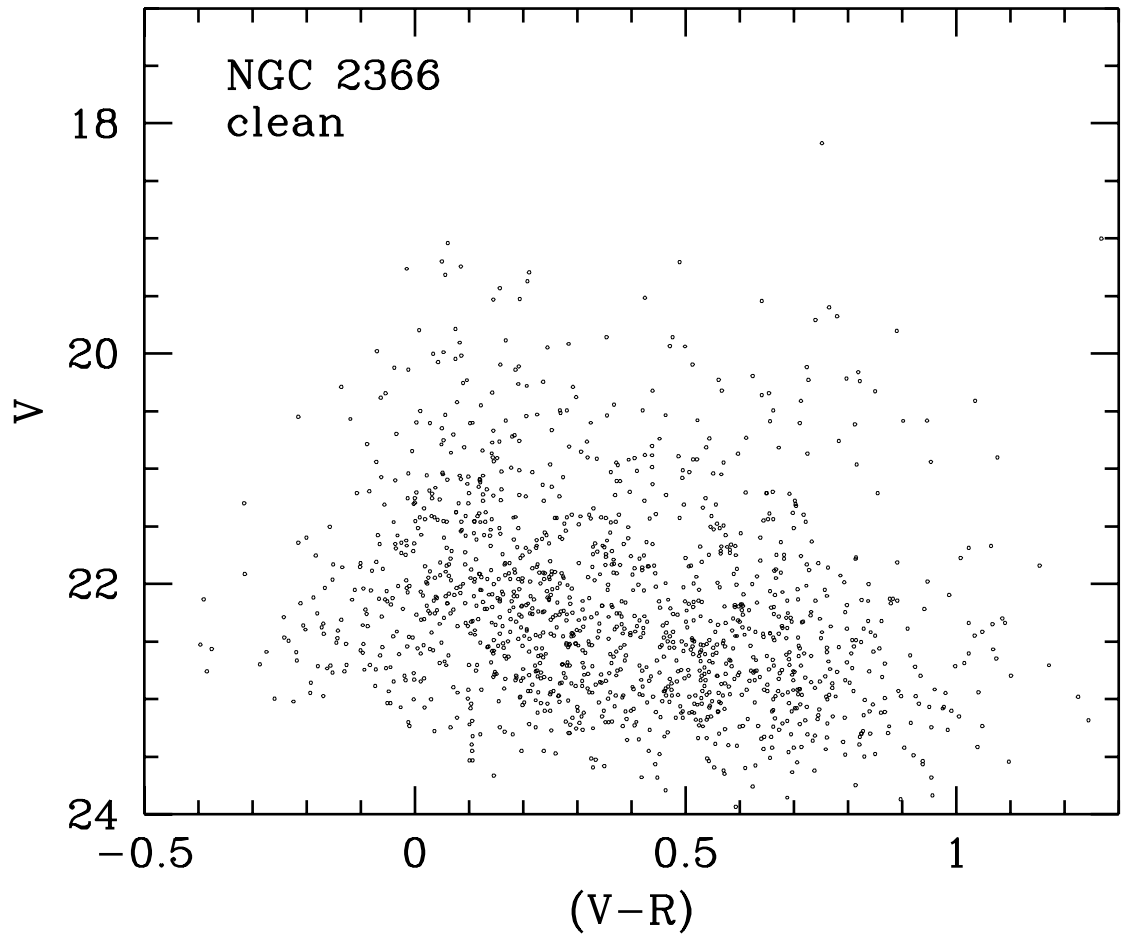

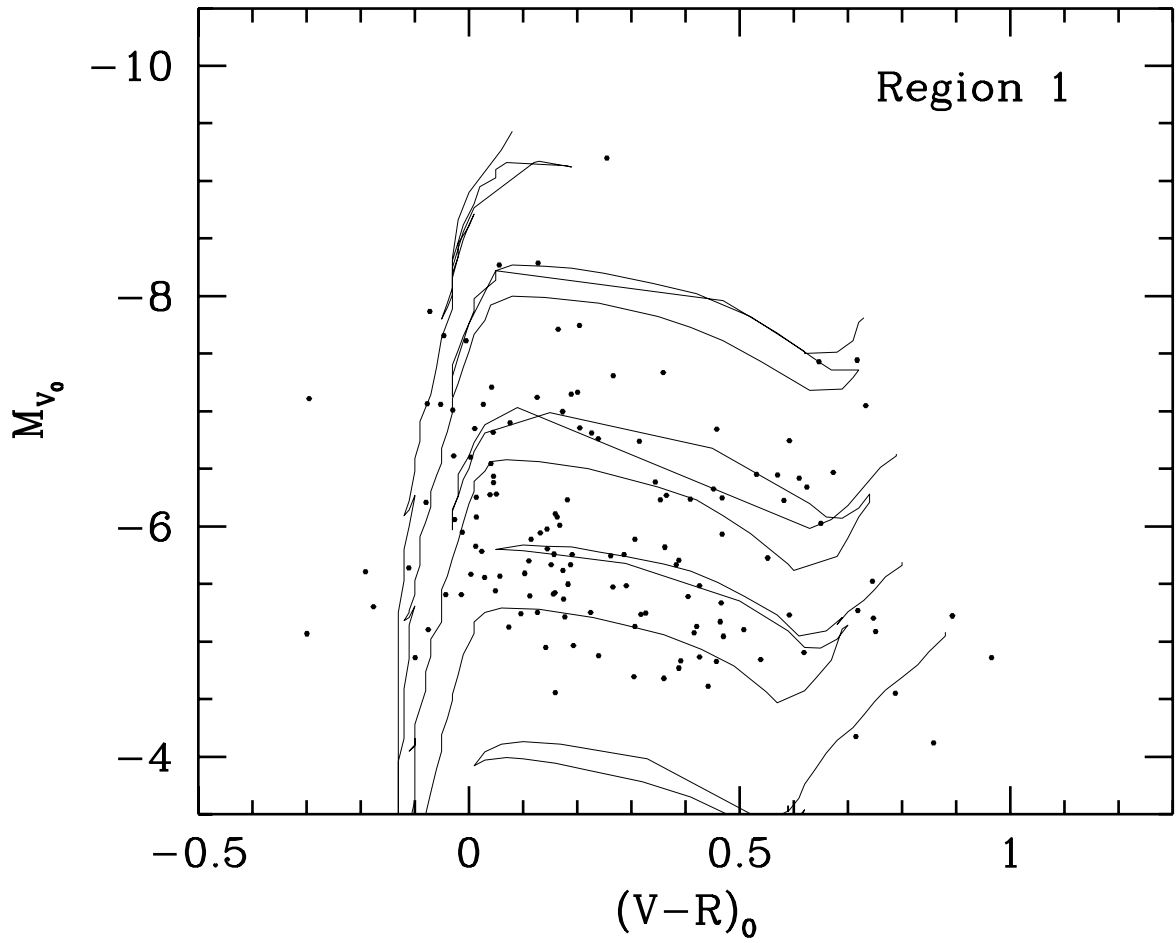

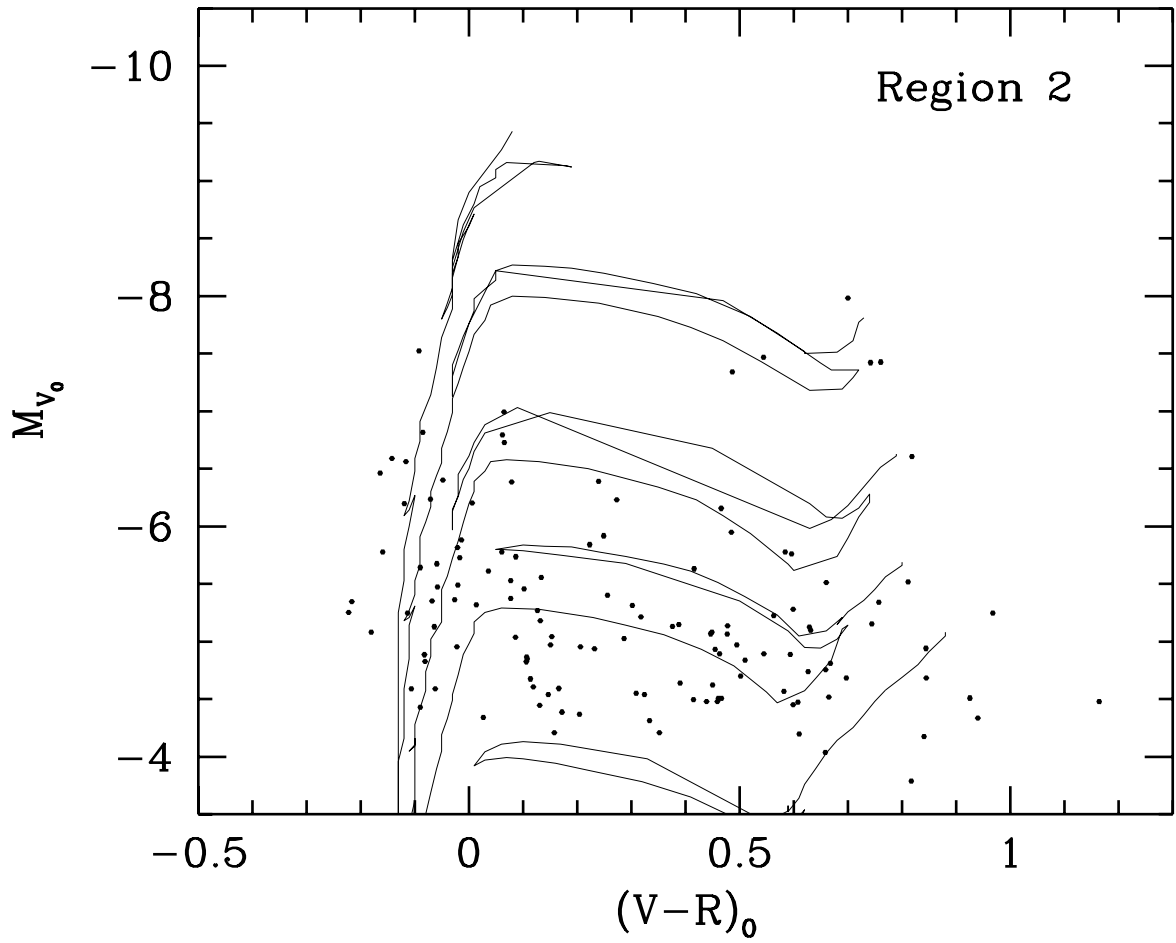

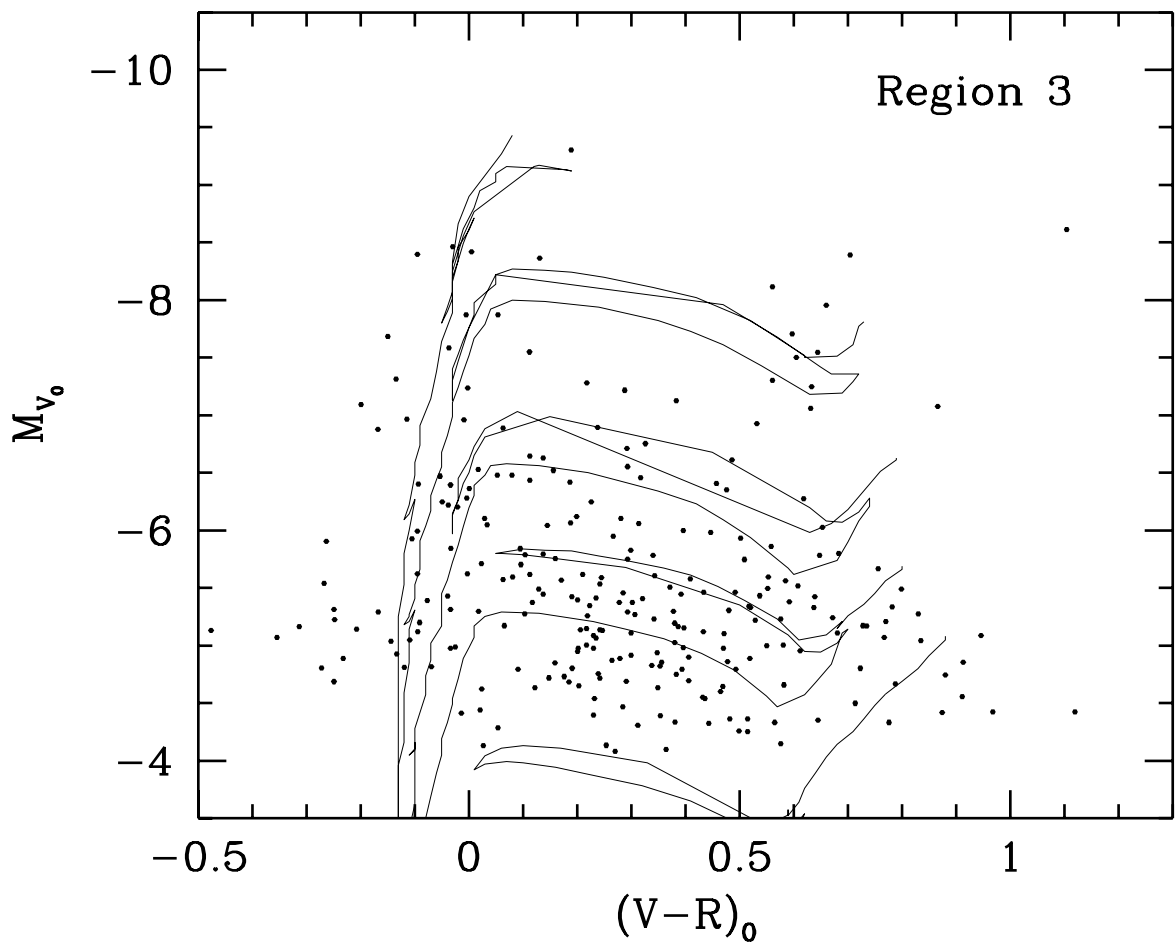

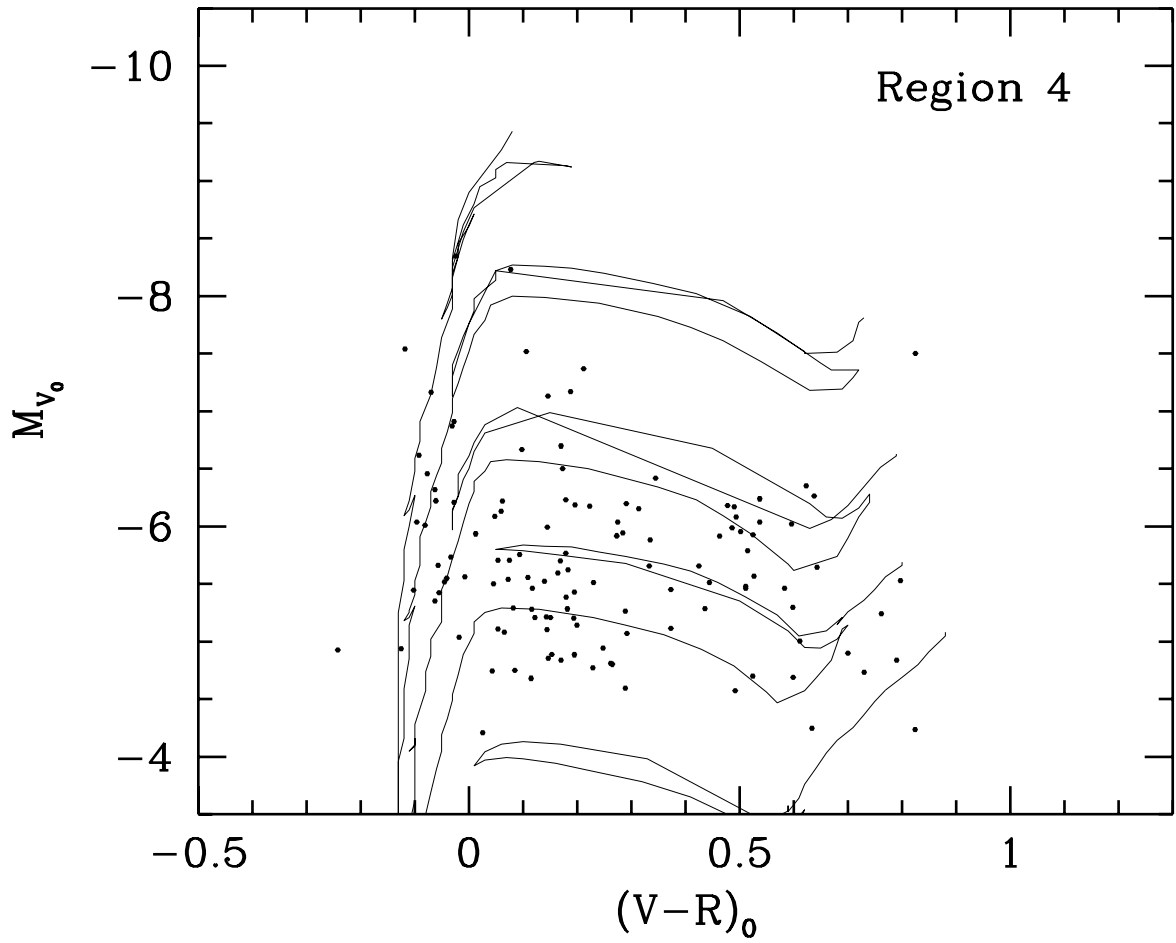

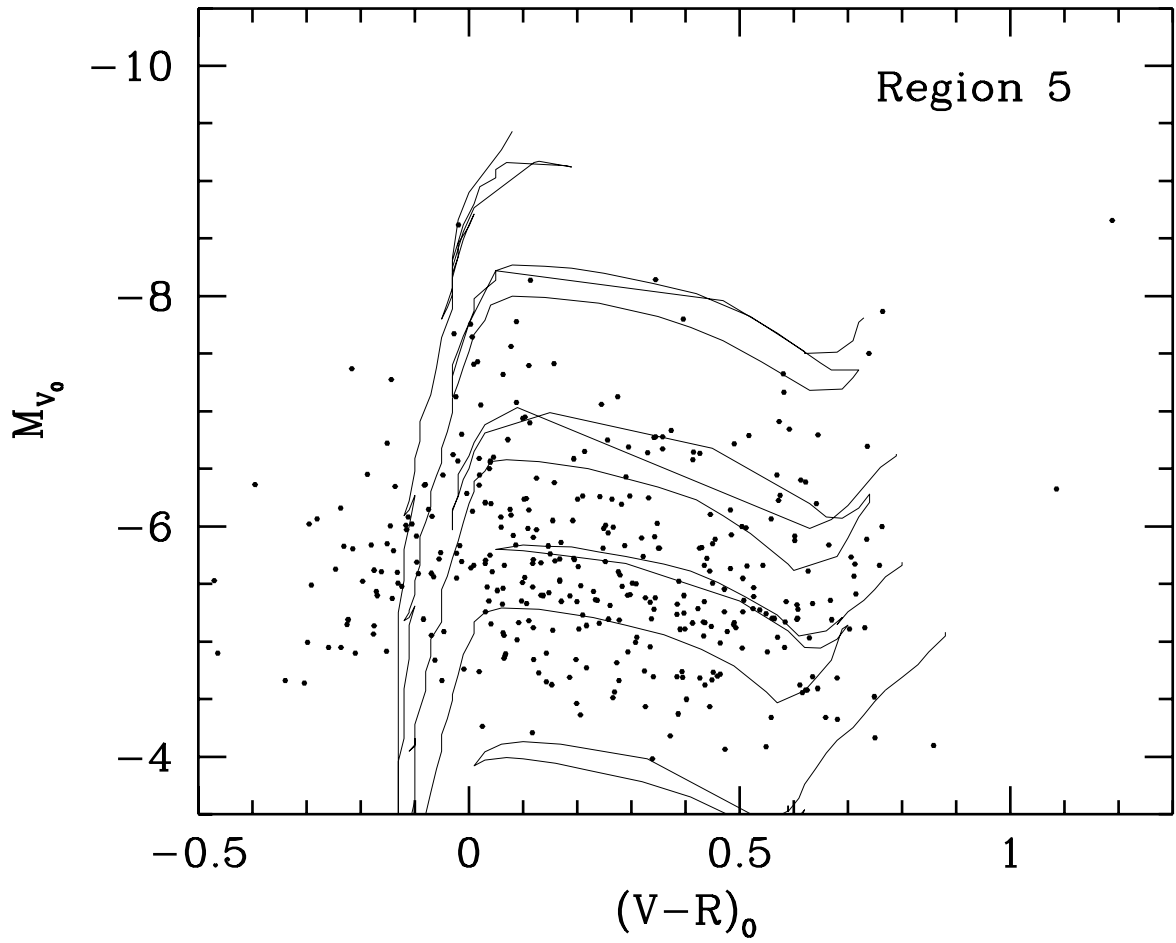

## Table 1. Global Parameters for NGC 2366

| | | |
|---:|:---:|:---:|
| $\alpha_{1950}$: | $07^h23^m37^s$ | |
| $\delta_{1950}$: | $+69°19'.1$ | |
| $l_{1950}$: | $146°.43$ | |
| $b_{1950}$: | $+28°.54$ | |
| $m_B^{bi}$: | $11.03 - 11.12$ | 1,2 |
| $m_V^{bi}$: | $10.62$ | 3 |
| $m_R^{bi}$: | $10.74$ | 1 |
| $m_I^{bi}$: | $10.82$ | 1 |
| $m_H^{bi}$: | $10.78$ | 1 |
| $A_B = A_B^b + A_B^i$: | $0.47 - 0.52$ | 1,3,4,5 |
| $i$: | $63° - 69°$ | 1,10 |
| $(m-M)_0$: | $27.07 - 27.65$ | 6,7,8,9 |
| $W_{20}$: | $113 - 118 Kms^{-1}$ | 1,10,11 |
| $W_{50}$: | $96 Kms^{-1}$ | 11 |
| $v_{LG}$: | $285 Kms^{-1}$ | 10 |
| $M_T$: | $(1.8 - 4.6) \times 10^9 M_\odot$ | 10,12 |
| $M_{HI}/M_T$: | $0.33 - 0.48$ | 10,12 |
| $M_T/L_B$: | $2.4$ | 8 |
| $12 + \log(O/H)$: | $7.81 - 8.02$ | 13,14,15 |

**Notes:** $\alpha_{1950}$, $\delta_{1950}$: equatorial coordinates; $l_{1950}$, $b_{1950}$: galactic coordinates; $m_B^{bi}$, $m_V^{bi}$, $m_R^{bi}$, $m_I^{bi}$, $m_H^{bi}$: apparent magnitudes corrected for galactic and internal extinction; $A_B$: total (galactic and internal) extinction in B; $i$: inclination angle; $(m-M)_0$: true distance modulus; $W_{20}$, $W_{50}$: line width of the 21 cm HI line measured respectively at 20% and 50% of peak intensity; $v_{LG}$: velocity relative to the Local Group barycenter, using the Yahil, Tammann & Sandage's (1977) correction for the solar motion; $M_T$: total mass; $M_{HI}/M_T$: Hydrogen to total mass ratio; $M_T/L_B$: mass-luminosity relation; $12 + \log(O/H)$: Oxygen abundance relative to Hydrogen.

**References:** (1) Pierce & Tully 1992; (2) Kraan-Korteweg et al 1988; (3) Shimasaku & Okamura 1992; (4) de Vaucouleurs et al 1991; (5) Burstein & Heiles 1984; (6) Tikhonov et al 1991; (7) Sandage & Tammann 1974; (8) de Vaucouleurs 1978b; (9) Tully 1987; (10) Hutchmeier & Richter 1986; (11) Bottinelli et al 1990; (12) Wevers et al 1986; (13) Masegosa et al 1991; (14) Skillman et al 1989; (15) González-Delgado et al 1994.

## Table 2. Journal of Observations

| Date | Time (UT) | Filter | Exp. time (s) | FWHM (") |
|---|---|---|---|---|
| March 21, 1992 | 20:33 | V | 2100 | 1.4 |
| March 21, 1992 | 21:22 | R | 1800 | 1.5 |
| March 21, 1992 | 22:13 | B | 3500 | 2.0 |

Table 4. Comparison with previous photoelectric sequences

| Star | $B_{ST}$ | $V_{ST}$ | $B_{TBKG}$ | $V_{TBKG}$ | $B_{here}$ | $V_{here}$ |
|------|----------|----------|------------|------------|------------|------------|
| B | 13.94 | 13.03 | – | – | 14.28 | – |
| C | 14.43 | 13.76 | 14.49 | 13.70 | 14.53 | – |
| D | 15.79 | 15.21 | – | – | 15.91 | 15.37 |
| E | 18.04 | 16.87 | 18.02 | 16.74 | 18.00 | 16.68 |
| F | 19.63 | 18.63 | 19.82(pg) | 18.74(pg) | 19.70 | 18.50 |
| G | 20.16 | 19.07 | 20.10(pg) | 19.03(pg) | 20.10 | 18.86 |
| H | 19.75 | 19.28 | 19.89(pg) | 19.18(pg) | 19.84 | 18.91 |
| I | 17.64 | 16.98 | 17.58(pg) | 16.92(pg) | 17.78 | 17.04 |

Table 5. Distance estimates to NGC 2366

| Estimator | $(m-M)_0$ | Reference |
|---|---|---|
| Blue supergiants | 27.07 | Sandage & Tammann (1974) |
| " | 27.10 | de Vaucouleurs (1978b) |
| " | 27.65 | Tikhonov et al (1991) |
| " | 27.00 | This paper |
| Red supergiants | 27.62 | Tikhonov et al (1991) |
| " | 27.60 | This paper |
| HII regions | 27.07 | Sandage & Tammann (1974) |
| Local velocity field | 27.35 | Tully (1987) |

Table 6. Integrated magnitudes and colors

| Zone | $V_0^{int}$ | $(B-V)_0^{int}$ | $(V-R)_0^{int}$ |
|------|-------------|------------------|------------------|
| 1    | 22.95       | 0.06             | 0.07             |
| 3    | 22.14       | 0.07             | 0.09             |
| 4    | 22.83       | 0.00             | −0.02            |
| 5    | 22.39       | 0.12             | 0.15             |